\documentclass[12pt]{article}

\usepackage{amsfonts, amsmath, amssymb}
\usepackage[dvips]{graphicx}
\usepackage{cite}

\newcommand{\bmat}{\left(\begin{array}}
\newcommand{\emat}{\end{array}\right)}

\newcommand{\pc}{\mathbb{C}}
\newcommand{\pr}{\mathbb{R}}

\def\ts{\textstyle}

\def\a{\alpha}

\def\b{\beta}
\def\g{\gamma}
\def\G{\Gamma}
\def\d{\delta}

\def\-{\hphantom{-}}

\def\s2{\frac{1}{\sqrt2}}

\def\beq{\begin{equation}}
\def\eeq{\end{equation}}
\def\beqa{\begin{eqnarray}}
\def\eeqa{\end{eqnarray}}

\def\T{{\rm T}}
\def\Z{{\mathbb Z}}

\def\cn{{\mathcal N}}

\def\deq#1{\mbox{$d$=#1}}
\def\neq#1{\mbox{$\cn$=#1}}
\def\Dsl{\,\raise.15ex\hbox{/}\mkern-13.5mu D} 
\def\r#1{\mbox{{\bf #1}}}

\def\r#1{{\bf #1}}

\def\vac{|0\rangle}

\begin{document}
\pagestyle{plain}

\makeatletter
\@addtoreset{equation}{section}
\makeatother
\renewcommand{\theequation}{\thesection.\arabic{equation}}
\pagestyle{empty}
\vspace{0.5cm}
\begin{center}
\LARGE{
 Heterotic strings on $G_2$ orbifolds
\\[10mm]}
\large{Anamar\'{\i}a Font \\[6mm]}
\small{
Departamento de F\'{\i}sica, Centro de F\'{\i}sica Te\'orica y Computacional \\[-0.3em]
Facultad de Ciencias, Universidad Central de Venezuela\\[-0.3em]
A.P. 20513, Caracas 1020-A, Venezuela
\\[8mm]} 
\small{\bf Abstract} \\[5mm]
\end{center}
\begin{center}
\begin{minipage}[h]{14.0cm} 
We study compactification of heterotic strings to three dimensions
on orbifolds of $G_2$ holonomy. We consider the standard embedding
and show that the gauge group is broken from $E_8 \times E_8^\prime$ or $SO(32)$
to $F_4 \times E_8^\prime$ or $SO(25)$ respectively. We also compute the spectrum of 
massless states and compare with the results obtained from reduction
of the 10-dimensional fields. Non-standard embeddings are discussed
briefly. For type II compactifications we verify that IIB and IIA have 
equal massless spectrum.

\end{minipage}
\end{center}
\newpage
\setcounter{page}{1}
\pagestyle{plain}
\renewcommand{\thefootnote}{\arabic{footnote}}
\setcounter{footnote}{0}

\section{Introduction}
\label{s:intro}

In this note we examine compactifications of type II and heterotic strings on a class of orbifolds
$\T^7/\Z_2^3$ whose singularities can be resolved to get manifolds with $G_2$ holonomy \cite{j1,j2,jbook}.
One motivation to undertake this problem is that, to our knowledge, hitherto it has received little attention. 
Compactification of the low-energy supergravity limits on compact 7-manifolds of $G_2$ holonomy
has been studied in \cite{papadopoulos} (see also \cite{ag}), but our
main concern is to discuss the string compactification from the world sheet perspective.

It is known that string compactification on singular orbifolds is well defined as long as
all twisted sectors are included and a projection on orbifold invariant states is implemented \cite{dhvw}.
We will precisely carry out this program to systematically construct the massless states and 
identify their multiplicities and gauge transformation properties. We will mainly focus on
heterotic strings with standard embedding of the orbifold action, but our methods can also be applied to 
analyze non-standard embeddings.

Various aspects of string compactification on $\T^7/\Z_2^3$ orbifolds have been investigated by several authors 
\cite{sv,acharya1,acharya,Majumder,Ferretti,gk,Barrett}. Compactification of type II strings on a different class of
compact $G_2$ manifolds was studied in \cite{Eguchi,bb}. Closer to our endeavor is the work \cite{sy} where it was shown
that in the standard embedding $E_8$ is broken to $F_4$. We rederive this result in the orbifold construction.

In the following we will first review the basic features of the $\T^7/\Z_2^3$ orbifolds, emphasizing the fixed point
structure and the implications for the orbifold partition function. In section \ref{s:type2} we consider type II
compactifications, both to obtain the full massless spectrum and to prepare the ground for the heterotic case.
In section \ref{s:het} we discuss compactification of the $SO(32)$ and $E_8\times E_8^\prime$ heterotic strings
in detail, including non-standard embeddings. In order to compare results we also describe compactification on
smooth 7-manifolds of $G_2$ holonomy. To this end for completeness in appendix 4 we construct the gravitino zero
modes which also give the gaugino zero modes that determine the number of charged multiplets.

\section{Joyce orbifolds}
\label{s:jorbi}

We will consider Joyce orbifolds of type $\T^7/\Gamma$ with automorphism group $\Gamma=\Z_2^3$.
The torus itself is a quotient $\pr^7/\Z^7$ and has coordinates $(x_1,\cdots, x_7)$, with
$x_i \equiv x_i + 1$. The generators of $\Gamma$, denoted $\a$, $\b$ and $\g$, are isometries of $\T^7$ 
that act on the coordinates as
\beqa
\a((x_1, \cdots, x_7)) & = & (-x_1,-x_2,-x_3,-x_4,x_5,x_6,x_7)  \nonumber \\[2mm]
\b((x_1, \cdots, x_7)) & = & (-x_1+b_1,-x_2+b_2,x_3,x_4,-x_5,-x_6,x_7)  \label{abg} \\[2mm]
\g((x_1, \cdots, x_7)) & = & (-x_1+c_1,x_2,-x_3+c_3,x_4,-x_5+c_5,x_6,-x_7)  \nonumber 
\eeqa
where $b_i$ and $c_i$ are shifts equal to 0 or $\frac12$. For concreteness we focus in two 
examples\footnote{These are examples 3 and 4 in section 3 of \cite{j2}.}, both having
$(b_1,b_2,c_1,c_5)=(0,\frac12,\frac12,0)$ but distinguished by whether $c_3=\frac12$ in model A,
or $c_3=0$ in model B. The shifts are appropriately chosen to ensure that after resolving the 
orbifold singularities the resulting manifold has $G_2$ holonomy \cite{j1,j2,jbook}.

A group element $\theta \in \Gamma$ acts as a rotation plus a translation, namely 
$\theta \vec{x} = \hat\theta \vec{x} + \vec{v}$,
where $\hat\theta \in SO(7)$. For example, $\hat\a={\rm diag}(-1,-1,-1,-1,1,1,1)$
and it is evident how it acts on vectors and tensors of $SO(7)$. It is also important to determine
the action on a spinor of $SO(7)$. Acting on the 8-dimensional spinor representation, the generator corresponding
to $\hat\theta$, denoted $P_\theta$, must satisfy   
\beq
P_\theta \, \Gamma^m \, P^{-1}_\theta = \hat\theta^m_{\; n}\, \Gamma^n
\label{pspin}
\eeq
where $\Gamma^m$ are Dirac matrices that fulfill $\{\G^m,\G^n\}=2\delta^{mn}$, $m,n=1,\cdots,7$.
It then follows that
\beq
P_\a = \G^1\G^2 \G^3 \G^4 \quad ; \quad  P_\b = \G^1\G^2 \G^5 \G^6 \quad ; \quad
P_\g = \G^1\G^3 \G^5 \G^7
\label{pgens}
\eeq
These matrices commute among themselves and therefore can be diagonalized 
simultaneously. There is only one common eigenvector with eigenvalues $(1,1,1)$ under $(P_\a, P_\b, P_\g)$. 
The remaining seven eigenvectors have eigenvalues that match those of $(\hat\a, \hat\b,\hat\g)$ acting on the $SO(7)$ 
vector\footnote{This can be checked using the 8-dimensional $\G$ matrices in section 8.2 of \cite{DNP}, multiplied by $-i$ 
to adjust conventions.}. 

According to the above discussion, the spinor $\r{8}$ of $SO(7)$ transforms
under $\Gamma$ as $\r1 + \r7$. This is precisely the decomposition of the $\r{8}$ under
$G_2 \supset SO(7)$, a first hint that the holonomy of the resolved $\T^7/\Gamma$ is $G_2$.
Furthermore, the group $\Gamma$ preserves the 3-form
\beq
\phi = dx_{127} + dx_{136} + dx_{145} + dx_{235} - dx_{246} + dx_{347} + dx_{567}
\label{phif}
\eeq
where $dx_{ijk} = dx_i \wedge dx_j \wedge dx_k$. The dual 4-form ${}^*\phi$ is also preserved.
These forms are defined by the structure constants of the octonion algebra, and the subgroup of $GL(7,\pr)$
that leaves them invariant is $G_2$. 

\subsection{The resolved orbifolds}
\label{ss:resolved}

We need to inspect in some detail the singularities of the orbifolds introduced above.
In the two models distinguished by $c_3$, the only elements besides the identity having fixed points are $\a$, $\b$
and $\g$ because other elements involve pure translations in $x_1$ or $x_2$. In fact, the fixed sets of each element
are 16 copies of $\T^3$. In both models, the group generated by $\b$ and $\g$ acts freely on the 
fixed points of $\a$, and similarly the fixed points of $\b$ are not left invariant by the sub-group
generated by $\a$ and $\g$. Instead the fixed points form orbits. For instance, the 16 fixed points
of $\a$ that have coordinates $(x_1,x_2,x_3,x_4)$ with $x_i=f_i=0, \frac12$, span the four orbits 
\beq
\big\{ (0,0,f_3,f_4) + (0,\textstyle{\frac12},f_3,f_4) + (\frac12,0,f_3,f_4) + (\frac12, \frac12, f_3,f_4) \big\}
\label{aorbit}
\eeq
Then, the singular set of $\a$ has four components with geometry
\beq
\T^3 \times \pc^2/\Z_2
\label{singalpha}
\eeq
where $\pc^2$ has coordinates $z_1=x_1+ix_2$, $z_2=x_3+ix_4$, and $\Z_2$ is the action $(z_1,z_2) \to
(-z_1,-z_2)$. Each singularity can be resolved by replacing it with an Eguchi-Hanson space \cite{j1,j2}.
The singular set of $\b$ is analogous.  

The behavior of the fixed points under $\g$ depends on the value of $c_3$. In model A with $c_3=\frac12$, the subgroup
generated by $\a$ and $\b$ acts freely on the fixed points of $\g$. In this case the singular set of $\g$ also consists 
of four components of the form (\ref{singalpha}). However, in model B with $c_3=0$, the element $\a\b$
leaves the fixed points of $\g$ invariant. Thus, the 16 fixed points of $\g$ organize into eight orbits, each
of two elements permuted by $\a$ and $\b$ but fixed by $\a\b$. They have coordinates  
 $(x_1,x_3,x_5,x_7)$ of the form 
\beq
\big\{ (\textstyle{\frac14},f_3,f_5,f_7) + (\textstyle{\frac34},f_3,f_5,f_7) \big\}
\label{gorbit}
\eeq
where $f_i=0, \frac12$ as before. In this orbifold the singular set of $\g$ has eight components described by
\beq
\big\{\T^3 \times \pc^2/\Z_2 \big\}/\Z_2^\prime
\label{singgamma}
\eeq
where now $\pc^2$ has coordinates $z_1=x_1+ix_7$, $z_2=x_3+ix_5$ and $\Z_2$ acts as before.
Including the $\T^3$ coordinates the action of the $\Z_2^\prime$ generator $\a\b$ is
\beq
\a\b : (z_1,z_2,x_2,x_4,x_6) \to (z_1,-z_2,x_2+\textstyle{\frac12},-x_4,-x_6) 
\label{abaction}
\eeq
Each singular component can be repaired by using an Eguchi-Hanson (EH) space but as explained in \cite{j2,jbook}, there
are two distinct ways of implementing the action of $\a\b$. In the option to blow up the singularity the orientation
of the EH space is preserved so that its fundamental 2-form $\omega_2$ is invariant under $\a\b$.
If the singularity is instead deformed, the orientation is reversed and $\omega_2$ picks up a minus sign under $\a\b$.

For future purposes it is useful to review the computation of the Betti numbers of the resolved $\T^7/\Z_2^3$ orbifolds.
By Poincar\'e duality $b^{7-k}=b^k$, $k=0,\cdots, 7$. A compact connected space of $G_2$ holonomy has $b^0=1$ and
$b^1=0$. Then, the only non-trivial Betti numbers are $b^2$ and $b^3$.
It is easy to see that the resolved orbifold indeed has $b^0=1$ and $b^1=0$, because on $\T^7$ there is one 0-form
but no 1-forms invariant under $\G$, and the resolution does not contribute to either $b^0$ or $b^1$. 
On $\T^7$ there are no invariant 2-forms either,
but there are seven $\G$ invariant 3-forms, namely the seven terms in $\phi$ in eq. (\ref{phif}). 
The four singular components of $\a$ are replaced by an EH space, each giving one 2-form $\omega_2$ and
three 3-forms $\omega_2 \wedge dx_i$, $i=5,6,7$. In model A the fixed points of $\b$ and $\g$
are repaired in the same manner. Then, the Betti numbers of the resolved orbifold, denoted, $Y_{\rm A}$ are
$b^2(Y_{\rm A})=12$ and $b^3(Y_{\rm A})=43$. 

In model B in which the singularities of $\g$ are eight copies of (\ref{singgamma}), the Betti
numbers $b^2$ and $b^3$ depend on how the $\Z_2^\prime$ acts on the EH space. We will consider only the case
in which all singularities are resolved in the same way. If the singularities are blown up
the 2-form $\omega_2$ of each EH space is invariant under
$\a\b$ and there will also be one additional invariant 3-form $\omega_2\wedge dx_2$.
In this case the Betti numbers of the resolved orbifold $Y_{\rm B1}$ are $b^2(Y_{\rm B1})=16$ and $b^3(Y_{\rm B1})=39$.
Instead, when the singularities are deformed each EH adds two invariant 3-forms $\omega_2\wedge dx_4$ and $\omega_2\wedge dx_6$.
In this situation the Betti numbers turn out to be $b^2(Y_{\rm B2})=8$ and $b^3(Y_{\rm B2})=47$.
Notice that in both B examples, as well as in model A, the sum of $b^2$ and $b^3$ is 55.

\subsection{Partition function}
\label{ss:pf}

The structure of the singular sets can be translated into properties of the partition function of strings
propagating on the $\T^7/\G$ orbifold. Recall that when $\G$ is Abelian this partition function 
can be written as \cite{dhvw}
\beq
Z = \sum_{h \in \G} Z_h=  \frac{1}{|\G|} \sum_{g,h \in \G} Z(h,g)
\label{pfunction}
\eeq
The sum over $h$ is over twisted sectors and the sum over $g$ enforces the orbifold projection. Correspondingly,
$Z(h,g)$ is the trace over states evaluated with boundary conditions periodic up to $h$ in the spatial direction
and up to $g$ in the time direction of the world-sheet torus. $Z_h$ is called the $h$-sector partition function.

In the $\T^7/\G$ orbifolds under study the partition function greatly simplifies because
$Z(h,g)$ vanishes when $h,g \in \G$ do not have simultaneous fixed points.
Moreover, for the $\G=\Z_2^3$ that we are considering the only sectors where massless
states can appear have $h=1, \a, \b, \g$.
In the remaining twisted sectors the lowest lying states are massive because $h$ acts as a pure translation
on some coordinates in which the winding numbers must then be half-integers.

When $h=1$, $g$ can be any element so that in the untwisted sector partition function the sum is over
all $g \in \G$.
On the other hand, in the $\a$-twisted sector with $h=\a$, $g$ can only be
the identity or $\a$ itself because other elements act freely on the $\a$ fixed points. Therefore, in this sector
\beq
Z_\a = \frac18\big[Z(\a,1) + Z(\a,\a)\big]
\label{pza}
\eeq
Since $\a$ leaves 16 fixed points, we see that in $Z_\a$ the states will appear with multiplicity four, consistent
with the fact that the singular set of $\a$ has four components. Notice that the orbifold projection just requires that
the states in the $\a$ sector be invariant under the subgroup generated by $\a$. 
In the untwisted sector the states must be invariant under the full $\G$.

The contribution $Z_\b$ of the $\b$ sector is analogous, and also $Z_\g$ in model A. 
In model B, the element $\a\b$ leaves the fixed points of $\g$ invariant so that
\beq
Z^{\rm B}_\g = \frac18\big[Z(\g,1) + Z(\g,\g)+ Z(\g,\a\b) + Z(\g,\a\b\g) \big]
\label{pzg}
\eeq 
States will appear with multiplicty 8 and they must be invariant under the subgroup generated by
$\g$ and $\a\b$.

In the next sections we will study strings propagating in the Joyce orbifolds described above.
We first consider type II compactification as a warm up exercise and then turn
to the most interesting case of heterotic compactifications. The properties of the partition function 
will be essential to obtain the spectrum of massless states which will be basically determined by
the Betti numbers of the resolved orbifolds.

\section{Type II compactifications}
\label{s:type2}

To begin we quickly review the reduction of the type II supergravities on smooth manifolds of
$G_2$ holonomy \cite{papadopoulos}. The resulting theory has four supercharges which means \neq2 supersymmetry
in \deq3. To count the number of massless multiplets it is enough to look at bosonic zero modes, taking
into account that in \deq3 a vector is dual to a scalar, and that the \neq2 scalar multiplet
has a complex scalar. Reduction of the 10-dimensional NSNS fields (metric, 2-form and dilaton)
to \deq3 gives rise, on shell, to a dilaton, $b^2$ real scalars from the 2-form, and real metric moduli whose number is 
$b^3$ for a manifold of $G_2$ holonomy \cite{gpp}. For type IIA the RR 1-form and 3-form
reduce to one plus $b^2$ vectors, and $b^3$ real scalars. Altogether there are $(1+b^2+b^3)$ scalar
multiplets. For type IIB, reduction of the RR even forms, with self-dual 4-form, also leads to
$(1+b^2+b^3)$ \neq2 scalar multiplets in \deq3. In \cite{papadopoulos} it was conjectured that IIA and IIB
strings compactified on a manifold of $G_2$ holonomy are equivalent.
We will show that type IIB and type IIA strings compactified on
$\T^7/\Z_2^3$ orbifolds of $G_2$ holonomy have equal massless spectrum as expected.

\subsection{Orbifold compactification}
\label{ss:type2orb}

The goal is to deduce the massless spectrum from compactification of the world sheet degrees of freedom on the orbifold.
To this end we use the light cone NSR formulation and denote the left and right moving oscillators respectively
by $(\a^m_r, \psi^m_s)$  and $(\tilde\a^m_r, \tilde\psi^m_s)$, $m=1,\cdots,8$. 
The mode numbers, which depend on the sector, are all integers
or half-integers in the $\Z_2^3$ orbifolds under analysis. In the untwisted sector the mass formulas 
for the Neveu-Schwarz (NS) and Ramond (R) states are as usual, namely
\beqa
{\rm NS} &:& M^2= N_B + N_F -\ts{\frac12} , 
\nonumber\\[2mm]
{\rm R} &:& M^2= N_B + N_F ,
\label{umass}
\eeqa
where $N_B$ and $N_F$ are bosonic and fermionic occupation numbers. These formulas apply to both left and right movers.
The massless states are the $\r{8}_v$
$\psi^m_{-\frac12}\vac$ and the  $\r{8}_s$ $|s^a\rangle$ which survive the GSO projection enforced
by keeping states with $e^{i\pi F}=1$. Here $\vac$ is the NS ground state, whereas the $|s^a\rangle$ Ramond states 
are built from the Clifford
vacuum of the zero mode algebra $\{\psi_0^m,\psi_0^n\}=2 \delta^{mn}$. For right movers the results are analogous.
For definiteness we focus on type IIB in which the right moving GSO projection is $e^{i\pi\tilde F}=1$, both for 
the NS and R sectors. Type IIA will be discussed afterwards.

We next implement the orbifold projection on the untwisted states. The $\Z_2^3$ action is generated by the $SO(7)$
rotations $\hat\a$,$\hat\b$, and $\hat\g$ under which the NS states transform as $\r{8}_v=\r1 + \r{7}_v$.
Clearly, the singlet is $\psi^8_{-\frac12}\vac$ whereas the $\r{7}_v$ are the states 
$\psi^i_{-\frac12}\vac$, $i=1,\cdots,7$. The invariant NSNS states are thus the dilaton 
$\tilde \psi^8_{-\frac12}\vac \otimes \psi^8_{-\frac12}\vac$, plus seven moduli
$\tilde \psi^i_{-\frac12}\vac \otimes \psi^i_{-\frac12}\vac$.
The Ramond states $|s^a\rangle$ transform as an  $\r{8}$ spinor of $SO(7)$, which under $\Z_2^3$ also splits as
$\r1 + \r{7}_v$, as we explained in the previous chapter. 
We will denote $|s^0\rangle$ the singlet state and $|s^i\rangle$ the remaining states transforming as $\r{7}_v$.
For right movers we make the analogous decomposition.
The invariant RR states are therefore the axion 
$|\tilde s^0\rangle \otimes |s^0\rangle$, and seven scalar moduli  
$|\tilde s^i\rangle \otimes |s^i\rangle$. 

Altogether the invariant untwisted states comprise 
one axiodilaton multiplet plus seven additional scalar multiplets. The states from the NSNS and RR sectors
combine into complex scalars while the NSR and RNS sectors provide the fermionic partners.
This result is consistent with the calculation of the spectrum by reduction of the 10-dimensional fields.
We have seen that in general, besides the axiodilaton multiplet, there are $(b^2+b^3)$ scalar multiplets.
We also know that the untwisted sector corresponds to compactification on the unresolved $\T^7/\Gamma$, and
in $\T^7$ there are no invariant 2-forms and 7 invariant 3-forms. Hence we indeed expect 7 additional scalar
multiplets in the untwisted sector from $b^2_{\rm unt}=0$ and $b^3_{\rm unt}=7$. 

We now analyze the twisted sectors. The twisted boundary conditions have the effect of changing the zero point energy
and the mode numbers of the oscillators. Since in all twisted sectors the $SO(7)$ rotations have four -1 eigenvalues 
the mass formulas become\footnote{\label{zpoint}Recall that the zero point energy of a real boson 
is $-\frac{1}{24} + \frac14 \delta(1-\delta)$, with $\delta=0$ 
for periodic and $\delta=\frac12$ for antiperiodic boundary conditions. For fermions there is an overall minus sign.}
\beqa
{\rm NS} &:& M^2= N_B + N_F ,
\nonumber\\[2mm]
{\rm R} &:& M^2= N_B + N_F .
\label{tmass}
\eeqa
There are massless states because in NS, as well as in R, there are zero modes.

To be concrete we specialize to the $\a$ sector. In the following it is crucial to remember that
the partition function in the $\a$ sector (\ref{pza}) just involves a projection 
into $\hat\a$ invariant states so that the action of $\hat \b$ and $\hat \g$
is irrelevant. This enables us to bosonize the fermions $\psi^i$, $i=1,2,3,4$, and $\psi^j$, $j=5,6,7,8$, respectively
into complex bosons $(H_1,H_2)$ transforming under $\hat \a$, and $(H_3,H_4)$ invariant under $\hat \a$.
In the NS sector the zero modes are $\psi^i_0$, $i=1,2,3,4$, and massless 
states can be labelled by a $SO(4)$ spinor weight $(w_1,w_2)$, with $w_1,w_2=\pm\frac12$. The GSO projection
selects $w_1=w_2$ and the two surviving weights make up the $(\frac12,0)$ representation of $SO(4)$.
The corresponding states, denoted $|\pm, \pm\rangle$, are invariant under $\hat\a$ that acts by multiplication by a phase 
$e^{i\pi(w_1-w_2)}$. In the R sector the zero modes are $\psi^j_0$, \mbox{$j=5,6,7,8$,} and
the massless states can be labelled by an $SO(4)^\prime$ weight $(w_3,w_4)$, with $w_3,w_4=\pm\frac12$.
The GSO projection picks $w_3=w_4$ and the states, denoted $|\pm, \pm\rangle^\prime$, are trivially invariant under $\hat\a$.
For right movers the NS and R states are analogous.
Requiring invariance under $\hat\a$ leaves four invariant NSNS 
states $|\pm, \pm\rangle\otimes |\pm, \pm\rangle$,  and similarly
four invariant RR states $|\pm, \pm\rangle^\prime \otimes |\pm, \pm\rangle^\prime$, all being real scalars.
The partition function (\ref{pza}) implies that in the $\a$ sector
there is an overall multiplicity of four, corresponding to the four orbits of fixed points.
Therefore, in the $\a$ sector there are altogether 16 massless scalar multiplets, with fermionic partners arising in 
NSR and RNS sectors.
This means that the contribution to the Betti numbers from the $\a$ sector is $(b^2_\a + b^3_\a)=16$, in agreement
with the resolution of the four singular sets of $\a$ that adds one invariant 2-form and 3 invariant 3-forms
per set. 

We now work out the $\g$ sector in model B, which is different because the orbifold projection requires
invariance under the subgroup generated by $\hat\g$ and $\hat\a \hat \b$. The massless states in the left NS sector
arise from the zero modes $\psi_0^k$, $k=1,3,5,7$. It is convenient to bosonize $(\psi^1,\psi^7)$ and  
$(\psi^3,\psi^5)$ into $H_{(17)}$ and $H_{(35)}$, so that the massless states are labelled by weights
$(w_{(17)}, w_{(35)})=(\pm \frac12, \pm \frac12)$. The GSO projection imposes $w_{(17)}=w_{(35)}$ and the states
are invariant under $\hat\g$ that acts by multiplication by $e^{i\pi(w_{(17)}-w_{(35)})}$.
Under $\hat\a\hat\b$ the states are not invariant, they instead acquire a phase $e^{i\pi w_{(35)}}$. 
Hence, in the NSNS sector there are only two invariant states because it must be that the left and right components
of the weights verify $w_{(35)}=\tilde w_{(35)}$. The R sector is similar. The fermions
$\psi^\ell$, $\ell=2,4,6,8$, are bosonized into $H_{(28)}$ and $H_{(46)}$ and the massless states are labelled by weights
$(w_{(28)}, w_{(46)})=(\pm \frac12, \pm \frac12)$ with GSO projection $w_{(28)}=w_{(46)}$. The states
are invariant under $\hat\g$ but are multiplied by $e^{i\pi w_{(46)}}$ under $\hat\a\hat\b$. 
Thus, in the RR sector there are also only two invariant states.
Taking into account the overall multiplicity of 8 in the $\g$ sector gives 16 massless scalar multiplets.
The corresponding Betti numbers from the $\g$ sector satisfy $(b^2_\g + b^3_\g)=16$, again matching
the resolution that adds two invariant 3-forms or one invariant 2-form plus one invariant 3-form per each of the 
eight singular sets of $\g$.

In model A the $\b$ and $\g$ twisted sectors are completely analogous to the $\a$ sector.
In model B the $\g$ sector is different but the net contribution is the same.
Including the untwisted states the full massless spectrum in both examples consists of 56 scalar multiplets.

So far we have discussed type IIB. For type IIA we just have to change the GSO projection for the right movers
in the R sector. It is easy to see that the massless spectrum remains unaltered. In the untwisted sector the
spinor $\r{8}_c$ states also transform as $1 + \r{7}_v$ of $SO(7)$. In the $\a$ sector the GSO
projection gives a $(0,\frac12)^\prime$ of $SO(4)^\prime$ in the right R sector, but the four
RR states $|\pm, \pm\rangle^\prime \otimes |\pm, \mp\rangle^\prime$ are still invariant.
The $\b$ sector is similar, also the $\g$ sector in model A.
In the $\g$ sector in model B the only difference is $\tilde w_{(28)}=-\tilde w_{(46)}=\pm \frac12$,
while $w_{(28)}=w_{(46)}=\pm \frac12$, but there are anyway 2 invariant states with $w_{(46)}=\tilde w_{(46)}$.

\section{Heterotic compactifications}
\label{s:het}

In this section we first discuss reduction of ten dimensional heterotic supergravity on a smooth manifold of $G_2$ holonomy. 
This problem has been addressed in \cite{papadopoulos} and \cite{acharya} but assuming a gauge background that only leaves 
unbroken the maximal Abelian subgroup $U(1)^{16}$. In this paper we rather want to consider the standard embedding in which 
the gauge and the spin connection are equal. This problem was already investigated 
in \cite{sy} where the authors give the full modular invariant partition function 
of the $E_8\times E_8^\prime$ heterotic string compactified on a $\T^7/\Z_2^3$ orbifold. They also argue that
$E_8$ is broken to $F_4$, and count the massless states in the \r{26} representation using properties of the underlying 
conformal field theory. We will determine the massless spectrum using simpler standard orbifold techniques
\cite{imnq, pol} that can also be implemented to study other consistent embeddings.

\subsection{Reduction}
\label{ss:hetred}

Upon compactification the resulting theory has two supercharges, meaning \neq1 in \deq3. There will be neutral and
charged matter in scalar multiplets that have one real scalar. There will also appear gauge multiplets with one
real vector. In ten dimensions there is a gravity multiplet plus a Yang-Mills multiplet with gauge group
$E_8 \times E_8^\prime$ or $SO(32)$. Reduction of the gravity multiplet gives $(1+b^2+b^3)$ neutral scalar multiplets whose
bosonic fields are the dilaton, the zero modes of the 2-form, and the metric moduli. The fermionic partners come from
the dilatino and the components $\Psi_i$, $i=1, \cdots, 7$, of the gravitino. Notice in particular
that by supersymmetry the $\Psi_i$ must have $(b^2 + b^3)$ zero modes. These zero modes are constructed explicitly in
the appendix.

The Yang-Mills fields give rise in \deq 3 to a gauge multiplet of the unbroken group plus charged scalar
multiplets. Clearly, the bosons arise respectively from zero modes of $A_\mu^J$ and $A_i^J$, where $J$ is a gauge
index. To determine the resulting group and matter representations we need to specify the gauge background.
In the standard embedding the gauge connection is equal to the spin connection which is a $G_2$ gauge field. In the
$E_8 \times E_8^\prime$ this is embedded in $E_8$. The commutant of $G_2$ in $E_8$ is $F_4$ and to arrive at the
corresponding branching it is useful to consider first the adjoint decomposition under $E_8 \supset SO(9) \times SO(7)$
given by
\beq
\r{248} = (\r{36}, \r1) + (\r{16}, \r8) + (\r{9}, \r7) + (\r1,\r{21})
\label{br97}
\eeq            
Now, under $SO(7) \supset G_2$, $\r8=\r1 + \r7$ and $\r{21}=\r7 + \r{14}$, whereas
under $F_4 \supset SO(9)$, $\r{52}=\r{36} + \r{16}$ and $\r{26}=\r1 + \r9 +  \r{16}$. Then, under
$E_8 \supset F_4 \times G_2$ the adjoint branching becomes
\beq
\r{248} = (\r{52}, \r1) + (\r{26}, \r7) + (\r1,\r{14})
\label{br42}
\eeq 
To look for zero modes it is actually easier to analyze the gauginos $\chi^J$ in the various representations 
of $F_4 \times G_2$. For $G_2$ singlets, i.e. $J \in (\r{52}, \r1)$, the gauginos just satisfy the Dirac
equation and we know that there is one solution, namely the covariantly constant spinor. This is the usual
argument that explains the existence of massless gauge multiplets in the adjoint of the unbroken group.
When $J \in (\r{26}, \r7)$, the gauginos transform in the fundamental of $G_2$, equivalently of $SO(7)$, so that
they satisfy the same equation as the gravitinos $\Psi_i$. 
In the appendix we show that the $\Psi_i$ have $(b^2+b^3)$ zero modes, hence
there will be $(b^2+b^3)$ massless scalar multiplets transforming in the $\r{26}$ of $F_4$.
Finally, for the gauginos in the adjoint of $G_2$, $J \in (\r1, \r{14})$, we can only say that in general
there will be zero modes that give massless multiplets singlets under $F_4$. Clearly all these fields are neutral under
the hidden $E_8^\prime$ that just leads to an adjoint gauge multiplet in \deq 3.

For the $SO(32)$ heterotic string, the gauge group is broken to $SO(25)$. The adjoint decomposition under  $SO(25) \times G_2$
is given by
\beq
\r{496} = (\r{300}, \r1) + (\r{25}, \r7) + (\r1, \r7) + (\r1,\r{14})
\label{br252}
\eeq  
In this case there will be $(b^2+b^3)$ massless scalar multiplets transforming as $\r{25} + \r1$ of $SO(25)$, plus 
a number of additional singlets that is not a topological invariant.

\subsection{Orbifold compactification}
\label{ss:hetorb}

Our purpose is to derive the massless spectrum from compactification of the world sheet degrees of freedom.
For right movers we again use the light cone NSR formulation in which oscillators are denoted
$(\tilde\a^m_r, \tilde\psi^m_s)$, $m=1,\cdots,8$. The right massless states have been derived in the previous section,
in fact, the right mass formulas are exactly as in type II, c.f. (\ref{umass}) and (\ref{tmass}).
We will only look into bosonic states so that we just need to consider right movers in the $\widetilde{\rm NS}$ sector.
The $\widetilde{\rm R}$ sector leads to fermionic partners that complete full \neq1 supermultiplets in \deq3. 

The left movers include 8 real bosons and 32 real `gauge' fermions $\lambda^A$.
In the $SO(32)$ heterotic string all fermions belong to one set with GSO projection $e^{i\pi F}=1$. 
In the $E_8\times E_8^\prime$ heterotic string the fermions are split into two sets $\lambda^a$ and 
$\lambda^{\prime a}$, $a=1, \cdots, 16$,  for which there are separate 
NS and R boundary conditions and GSO projections $e^{i\pi F}=1$, $e^{i\pi F^\prime}=1$. 
Conventions are those of \cite{pol}. To perform 
the compactification it is necessary to specify how the orbifold generators act on the gauge fermions. We will mostly
focus on the standard embedding which automatically satisfies the level-matching condition required by modular invariance. 
For a $\T^7/\Z_2^3$ heterotic orbifold the standard embedding consists of choosing a subset 
$\lambda^i$, $i=1, \cdots, 7$, on which the $SO(7)$ rotations $(\hat\a, \hat\b, \hat\g)$ act in the same way as 
on the $\tilde\psi^i$. In section \ref{sss:nse} we describe other 
possible actions consistent with level-matching.  We work with the fermionic formulation because
the form of the orbifold generators precludes combining the $\lambda$'s into complex fermions that could be bosonized. 
For clarity of presentation in the following we study the two heterotic theories in order.

\subsubsection{Standard embedding in $SO(32)$ heterotic}
\label{sss:so}

Massless states can only arise in the NS sector of the left fermions in which the mass formula is given by
\beq
{\rm NS} \ : \ M^2= N_B + N_F + \Delta_{\rm NS}, 
\label{lns}
\eeq
where in the untwisted and twisted sectors $\Delta_{\rm NS}=-1$ and $\Delta_{\rm NS}=-\frac12$ respectively, 
as can be readily checked using the results in footnote \ref{zpoint}. In the untwisted sector the GSO projection 
and level matching at zero mass only allow left NS states either with one bosonic 
or two fermionic oscillators. Then, the untwisted $(\widetilde{\rm NS}, \rm{NS})$ massless states are
\beqa
\tilde\psi_{-\frac12}^8 \vac & \otimes &  \left\{\begin{array}{l}
\a_{-1}^8 \vac \\[1mm]
\lambda_{-\frac12}^I  \lambda_{-\frac12}^J \vac 
\end{array}
\right.
\label{nsu1} \\[2mm]
\tilde\psi_{-\frac12}^i \vac & \otimes &  \left\{\begin{array}{l}
\a_{-1}^i \vac \\[1mm]
\lambda_{-\frac12}^i  \lambda_{-\frac12}^J \vac \\[1mm]
\lambda_{-\frac12}^j  \lambda_{-\frac12}^k \vac \quad ; \quad \phi_{ijk} \not=0 
\end{array}
\right.
\label{nsu2}
\eeqa
where $I,J=8, \cdots, 32$, and $i,j,k=1,\cdots, 7$. The states 
in (\ref{nsu1}) correspond to the dilaton and 300 gauge vectors (on-shell) that give the adjoint representation of $SO(25)$.
The states in (\ref{nsu2}) are 7 metric moduli, seven scalars transforming as \r{25} of $SO(25)$, 
and 21 gauge singlets. All states are invariant under the orbifold action. In particular the gauge singlets are 
invariant whenever $i,j,k$ are such that $\phi_{ijk}$ given in (\ref{phif})
is non-zero. We have obtained the spectrum expected from the smooth compactification. 
Indeed, given that $b^2_{\rm unt}=0$ and $b^3_{\rm unt}=7$,
there must be 7 metric moduli and seven scalars transforming as $(\r{25}+\r{1})$ of $SO(25)$. 
There are also 14 additional gauge singlets.

In the twisted sectors the gauge fermions $\lambda^i$ behave analogous to the $\psi^i$ so that we can use 
the results of section (\ref{ss:type2orb}).
We consider first the $\a$ sector in which the left vacuum is a $SO(4)$ spinor because there are 
zero modes $\lambda_0^m$, $m=1,2,3,4$. Taking into account the GSO and orbifold projections the massless states are found to be
\beq
|\pm, \pm \rangle  \otimes   \left\{\begin{array}{l}
\lambda_{-\frac12}^I  |\pm, \pm \rangle \\[1mm]
\lambda_{-\frac12}^\ell  |\pm, \pm \rangle \\[1mm] 
\a_{-\frac12}^m  |\pm, \mp \rangle
\end{array}
\right.
\label{nsa}
\eeq
where $\ell=5,6,7$. In the $\a$ sector there is an overall multiplicity of 4 due to 
the number of fixed orbits. Therefore, altogether matter comprises 16 scalars in the \r{25} of $SO(25)$ and $16\cdot 3$ 
gauge singlets. There are also $16\cdot 4$ singlets which presumably entail metric moduli
and blowing-up modes as in Calabi-Yau orbifolds with standard embedding \cite{Hamidi, pol}. However, although the 
states include bosonic oscillators acting on the vacuum as in the Calabi-Yau orbifolds, they are all gauge singlets. 
Hence, after resolving singularities the gauge group remains $SO(25)$, in agreement with the smooth 
compactification carried out before.

For model B we need to elucidate the $\g$ sector in which there are zero modes 
$\lambda_0^n$, $n=1,3,5,7$. The vacuum is given by $SO(4)$ spinors
also denoted $|\pm, \pm \rangle$ and $|\pm, \mp \rangle$. Now we have to impose 
invariance under $\hat\g$ and $\hat\a\hat\b$. The resulting massless states are
\beqa
|+, + \rangle  & \otimes &   \left\{\begin{array}{l}
\lambda_{-\frac12}^I  |+, + \rangle \\[1mm]
\lambda_{-\frac12}^2  |+, + \rangle \quad ; \quad \lambda_{-\frac12}^{4,6}  |-, - \rangle  \\[1mm] 
\a_{-\frac12}^{1,7}  |-, + \rangle \quad ; \quad  \a_{-\frac12}^{3,5}  |+, - \rangle
\end{array}
\right.
\label{nsg1} \\[2mm]
|-, - \rangle  & \otimes &   \left\{\begin{array}{l}
\lambda_{-\frac12}^I  |-, - \rangle \\[1mm]
\lambda_{-\frac12}^2  |-, - \rangle \quad ; \quad \lambda_{-\frac12}^{4,6}  |+, + \rangle  \\[1mm] 
\a_{-\frac12}^{1,7}  |+, - \rangle \quad ; \quad  \a_{-\frac12}^{3,5}  |-, + \rangle
\end{array}
\right.
\label{nsg2}
\eeqa
Clearly the spectrum has a different structure compared to that in the $\a$ sector. However, since the overall multiplicity
due to fixed sets is now 8, in total there are again 16 scalars in the \r{25} of $SO(25)$, $16\cdot 3$ gauge singlets 
and $16\cdot 4$ metric moduli plus blowing-up modes.

The massless spectrum in the twisted sectors also agrees with the smooth compactification. 
It happens that $(b^2_{\rm twi}+b^3_{\rm twi})=16$
and in fact in each sector we have found 16 scalars transforming as $(\r{25}+\r{1})$ of $SO(25)$ plus 32 additional gauge singlets.
Concerning the remaining $16\cdot 4$ singlets per sector, we expect that $16\cdot 3$ 
become massive upon blowing-up while 16 remain massless as
metric moduli.

In conclusion, the gauge group is broken to $SO(25)$ and there are $b^2+b^3=55$ multiplets transforming as $(\r{25}+\r{1})$. 
There are also 110 gauge bundle moduli, 14 from the untwisted sector and 32 from each twisted sector.
The remaining states are 55 moduli multiplets plus 144 additional 
singlets that presumably become massive after resolving the singularities.

\subsubsection{Standard embedding in $E_8 \times E_8^\prime$ heterotic}
\label{sss:e8}

There is now a $({\rm NS}, \rm{NS}^\prime)$ sector for the left fermions in which the mass formula is just (\ref{lns}) replacing
$N_F \to N_F + N_F^\prime$. It is easy to see that in the untwisted sector 
the massless states are as in (\ref{nsu1}) and (\ref{nsu2}) but with $I,J=8, \cdots, 16$.
Thus, there will be 36 gauge vectors that furnish the adjoint of $SO(9)$ and 7 scalars that transform as \r{9}. There are new
states $\tilde\psi_{-\frac12}^8 \vac  \otimes \lambda_{-\frac12}^{\prime a}  \lambda_{-\frac12}^{\prime b} \vac$ 
that are vectors in the adjoint of $SO(16)^\prime$. Recall that in the standard embedding the 
fermions $\lambda^{\prime a}$ are totally inert under the orbifold action.

The main new feature in the $E_8 \times E_8^\prime$ heterotic is the existence of massless states in mixed 
sectors of the left fermions in which the mass formula turns out to be
\beq
({\rm NS}, {\rm R}^\prime), ({\rm R}, {\rm NS}^\prime) \ : \ M^2= N_B + N_F + N_F^\prime
\label{lrns}
\eeq
for both untwisted and twisted sectors. $({\rm NS}, \rm{R}^\prime)$ only leads to gauge vectors that transform 
in the \r{128} of $SO(16)^\prime$ and complete the adjoint of $E_8^\prime$.
In the untwisted $({\rm R}, \rm{NS}^\prime)$ sector there are zero modes $\lambda_0^a$ and the vacuum is a $SO(16)$ spinor.
The GSO projection selects the \r{128} that under $SO(7)\times SO(9)$ transforms as $(\r{8},\r{16})$. We already know that the
\r{8} spinor of $SO(7)$ transforms as $\r{1}+\r{7}_v$ under the orbifold $\Z_2^3$. Therefore, massless states can be denoted
$|s^0, S\rangle$ and $|s^i, S\rangle$, $i=1,\cdots, 7$, where $S$ stands for the \r{16} spinor of $SO(9)$.
The orbifold invariant states in $(\widetilde{\rm NS}, {\rm R}, \rm{NS}^\prime)$ are then
\beq
\tilde\psi_{-\frac12}^8 \vac  \otimes |s^0, S\rangle \quad ; \quad  \tilde\psi_{-\frac12}^i \vac  \otimes |s^i, S\rangle
\label{rnsu}
\eeq
Combining with states from $(\widetilde{\rm NS}, {\rm NS}, \rm{NS}^\prime)$ gives 52 gauge vectors that provide the adjoint of
$F_4$. In fact, under $F_4 \supset SO(9)$, $\r{52} = \r{36} + \r{16}$.
We find also 7 scalars transforming in the fundamental \r{26} of $F_4$
decomposed as $(\r{1} + \r{9} + \r{16})$ under $SO(9)$. There remain 14 extra gauge singlets.

Consider now the $\a$ twisted sector. 
There are massless solutions of (\ref{lrns}) on account of the zero modes $\lambda_0^{5, \cdots, 16}$.
The vacuum is a $SO(12)$ spinor which is $\a$ invariant. The GSO projection selects the \r{32} that transforms as  $(\r{2},\r{16})$
under $SO(3)\times SO(9)$. We label the states as $|\sigma, S\rangle$ with $\sigma$ the \r{2} spinor of $SO(3)$.
The invariant states in $(\widetilde{\rm NS}, {\rm R}, \rm{NS}^\prime)$ are simply
\beq
|\pm, \pm \rangle  \otimes |\sigma, S\rangle
\label{rnsa}
\eeq
which correspond to 4 scalars transforming as \r{16} of $SO(9)$. The $(\widetilde{\rm NS}, {\rm NS}, \rm{NS}^\prime)$ massless
states are read from (\ref{nsa}). We see that among them there are 4 scalars transforming as $\r{1} + \r{9}$ of $SO(9)$.
Including the multiplicity 4 due to the fixed sets we conclude that
there are 16 scalars in a full \r{26} of $F_4$. There remain $16\cdot 2$ gauge singlets.

Presently we analyze the $\g$ sector in the B model. In $({\rm R}, \rm{NS}^\prime)$ there are zero modes $\lambda_0^{2,4,6}$ and
$\lambda_0^{8, \cdots, 16}$. After the GSO projection the vacuum is again a $SO(12)$ spinor \r{32} that decomposes as
$(\r{2},\r{16})$ of $SO(3)\times SO(9)$. To perform the orbifold projection we need to determine how the spinor \r{2}
of $SO(3)$ transforms under $\hat \a \hat \b$. The $SO(3)$ restricts to the coordinates $(x_2,x_4,x_6)$ on which
$\hat \a \hat \b$ acts as ${\rm diag}(1,-1,-1)$. In the spinor representation the action is $P_{\a\b}=\g^2 \g^3$ where
the three 2-dimensional Dirac matrices satisfy $\{\g^i,\g^j\}=\d^{ij}$. Clearly $P_{\a\b}$ has eigenvalues $\pm i$, the
corresponding eigenstates are called $\sigma_{\pm}$. 
Thus, the $\g$ invariant states in $(\widetilde{\rm NS}, {\rm R}, \rm{NS}^\prime)$ are 
\beq
|+, + \rangle  \otimes |\sigma_+, S\rangle \quad ; \quad |-,- \rangle  \otimes |\sigma_-, S\rangle 
\label{rnsg}
\eeq
where $S$ again stands for the \r{16} spinor of $SO(9)$. Combining with states in $(\widetilde{\rm NS}, {\rm NS}, \rm{NS}^\prime)$
given in (\ref{nsg1}) and (\ref{nsg2}), yields 2 scalars transforming in the \r{26} of $F_4$.
Since the fixed set multiplicity is 8 in the end the overall spectrum is the same as in the $\a$ sector studied before.

The final outcome is that the massless orbifold spectrum conforms with reduction on a smooth manifold. $E_8$ is broken to
$F_4$ and there are 55 multiplets transforming in the \r{26}. The counting of additional singlets is exactly as in the
$SO(32)$ heterotic. Similar results have been obtained in \cite{sy}.

\subsubsection{Non-standard embeddings}
\label{sss:nse}

The orbifold action on the right fermions $\tilde \psi^i$ is given by the $SO(7)$ rotations $(\hat\a, \hat\b, \hat\g)$
defined in (\ref{abg}). The embedding in the left fermions $\lambda^A$ is realized by gauge twists of order two
denoted $(A,B,C)$. These twists can be taken to be diagonal and are specified by strings of $(-1)$'s and 1's. For instance,
in the standard embedding
\beqa
A_0 & = & (-1^4,1^{12}; 1^{16}) \nonumber\\
B_0 & = & (-1^2,,1^2,-1^2,1^{10}; 1^{16}) \label{seabc}\\
C_0 & = & (-1,1,-1,1,-1,1,-1,1^9; 1^{16}) \nonumber
\eeqa
where $(\pm 1)^n$ stands for $(\pm 1)$ repeated $n$ times. The separation into two 
groups of 16 eigenvalues is meant to apply only to the $E_8\times E_8^\prime$ heterotic string.

To determine the level-matching constraints we take a generic twist with $t$ eigenvalues equal to $(-1)$. 
The zero point energy for all left
fermions in NS sector is easily found to be $\Delta_{\rm NS}=\frac{t-12}{16}$, 
while $\Delta_{\rm R}=\frac{20-t}{16}$ in the R sector.
On the other hand, for the right movers, $\tilde \Delta_{\rm NS} = \tilde \Delta_{\rm R} = 0$. Therefore, level-matching requires 
that $t$ take values $4,12,20,28$, so that $\Delta_{\rm NS}$ is a multiple of $\frac12$, 
as the occupation numbers in the twisted sectors.
In the $E_8\times E_8^\prime$ heterotic string the $(-1)$ eigenvalues can be distributed between 
the two sets of fermions but the number on each
set has to be a multiple of 4. This last condition guarantees that the zero point energy in mixed $({\rm R}, {\rm NS}^\prime)$, 
$({\rm NS}^\prime, {\rm R})$ sectors is a multiple of $\frac12$ as well. 
It can also be understood in the bosonic formulation in which a $\Z_2$ twist must correspond to a
shift vector $V$ such that $2V$ belongs to the $E_8\times E_8^\prime$ root lattice.

For a single twist some possibilities are equivalent, e.g. $t=12$ and $t=20$ in $SO(32)$, but all have to be taken into account to
obtain the allowed triplets $(A,B,C)$. Notice that the products $AB$, $BC$ and $AC$ must also satisfy the condition on the number
of negative eigenvalues. We will not attempt to classify all allowed embeddings. We will just give some simple examples to show how
the standard orbifold techniques can be applied to derive the spectrum.

The first example in the $SO(32)$ heterotic has twists $(A_1,B_0,C_0)$, with
\beq
A_1  =  (-1^4,1^3,(-1)^{8},1^{17}) 
\label{a1s}
\eeq
and $B_0, C_0$ as in (\ref{seabc}). We will describe the spectrum briefly, concentrating in the
differences with the standard embedding. The gauge group turns out to be $SO(9)\times SO(17)$. The less
evident $SO(9)$ gauge vectors arise from untwisted states 
$\tilde\psi_{-\frac12}^8 \vac \otimes \lambda_{-\frac12}^r  \lambda_{-\frac12}^s \vac$, $r,s,=4,8,\cdots,15$, which happen
to be invariant. The untwisted charged matter consists of $[6(\r{9},\r1) + 6(\r1,\r{17}) + (\r9,\r{17})]$. For instance,
the mixed states are $\tilde\psi_{-\frac12}^4 \vac \otimes \lambda_{-\frac12}^r  \lambda_{-\frac12}^L \vac$, $L=12,\cdots,32$.
In the $\a$ twisted sector there are massless states in the NS sector of the left fermions but now the zero point energy
vanishes. The zero modes are $\lambda_0^{1,2,3}, \lambda_0^r$. Then, altogether the massless charged states are
$16(\r{16},\r1)$, where \r{16} is the $SO(9)$ spinor. The charged states in the $\b$ and $\g$ sectors are basically
as in the standard embedding. In each case we find $16[(\r{9},\r1) + (\r1,\r{17})]$. 

A non-standard embedding in the $E_8\times E_8^\prime$ heterotic is obtained with twists $(\tilde A_1,B_0,C_0)$, where now
\beq
\tilde A_1  =  (-1^4,1^3,(-1)^{4}, 1^5;(-1)^4,1^{12}) 
\label{at1s}
\eeq
The eight additional $(-1)$ eigenvalues are distributed between the two factors to avoid reobtaining the $F_4\times E_8^\prime$
model. The gauge group is found to be $SO(9)\times E_7^\prime \times SU(2)^\prime$. The charged matter spectrum can be determined 
as in the previous examples. For instance, in the $\a$ sector there are $16(\r{16}, \r1^\prime, \r2^\prime)$.

\section{Final Comments}
\label{s:con}

The aim of this paper was to study the compactification of heterotic strings on $\T^7/\Z_2^3$ orbifolds.
Using systematic orbifold techniques we were able to find the invariant massless states in the untwisted
and twisted sectors. In the standard embedding the results match those obtained from reduction of the 10-dimensional 
theory on a smooth manifold of $G_2$ holonomy. Concretely, the gauge group $SO(32)$ or
$E_8 \times E_8^\prime$ is broken to $SO(25)$ or $F_4 \times E_8^\prime$ respectively. Furthermore, there are
$(b^2 + b^3)$ multiplets transforming in the fundamental of $F_4$ or $SO(25)$, as well as an equal number
of moduli multiplets.  The Betti numbers are those of the resolved orbifold. There are additional gauge bundle moduli whose number 
is also determined in the orbifold construction. Our methods naturally apply to investigate non-standard embeddings
and we have provided some examples.

We have also shown that type IIB and type IIA strings compactified on
$\T^7/\Z_2^3$ orbifolds of $G_2$ holonomy have equal massless spectrum consisting of
$(1+b^2+b^3)$ \neq2 scalar multiplets in \deq3.

Our main motivation was to study standard and non-standard heterotic compactifications to uncover the unbroken
gauge symmetries. The allowed Higgsing patterns in the resulting  \deq3 \neq1 supersymmetric gauge theories
could be determined. It would be of interest to compare with M-theory compactifications on 8-dimensional manifolds of
${\rm Spin}(7)$ holonomy to understand the enhancing to non-simply laced groups.

\vspace*{1cm}

{\bf Acknowledgments}\\
I am grateful to S. Theisen for useful explanations and remarks, to K. Stelle for a helpful hint, 
and to L. Ib\'a\~nez for comments on the manuscript. 
Thanks are due to the Max Planck Institute f\"ur Gravitationsphysik for hospitality at several
stages of this work. A research grant No. PI-03-007127-2008 from CDCH-UCV is acknowledged.


\appendix
\section{Gravitino zero modes}

By supersymmetry the gravitino $\Psi_p$, $p=1,\cdots, 7$, must have $(b^2+ b^3)$ zero modes.
This result can be shown using general properties of a manifold $Y$ of $G_2$ holonomy. On $Y$ there
is a covariantly constant spinor $\eta$ and in consequence there exist a covariantly constant 3-form $\varphi$
and a 4-form $\Phi={}^*\varphi$ given by \cite{gpp}
\beq
\varphi_{mnp} = i \bar\eta \Gamma_{mnp} \eta \quad ; \quad 
\Phi_{mnpq} = -\bar\eta \Gamma_{mnpq} \eta
\label{tfour}
\eeq
We use the conventions of \cite{House}.

Zero modes of $\Psi_p$ satisfy the Rarita-Schwinger equation
\beq
\Gamma^{mnp} D_n \Psi_p = 0
\label{rarita}
\eeq
In $Y$ we can construct the solutions using the covariantly constant spinor and the harmonic forms, as it is done in Calabi-Yau 
compactifications \cite{chsw}. For example, a natural Ansatz consists of
\beq
\Psi_p^{(0)} = \varphi_{prs} \, \Gamma^{rs} \eta
\label{psi0}
\eeq
It immediately follows that $\Psi_p^{(0)}$ satisfies (\ref{rarita}) because
$\varphi$ is covariantly constant. This mode has $\Gamma^p \Psi_p^{(0)} \not= 0$ and is related by supersymmetry
to the volume modulus of the metric. 

The remaining harmonic 3-forms, $a^{(\tau)}_{prs}$, $\tau=1, \cdots, b^3-1$, give rise to
\beq
\Psi_p^{(\tau)} = a^{(\tau)}_{prs} \, \Gamma^{rs} \eta
\label{psif}
\eeq
Using identities of antisymmetrized products of $\Gamma$ matrices \cite{gam} and the fact that the $a^{(\tau)}$
are closed and co-closed we find that $\Gamma^{mnp} D_n \Psi_p^{(\tau)} = \frac23 g^{mr} D_r(\Gamma^{nps} a_{nps}^{(\tau)} \eta)$.
Furthermore, $\Gamma^p \Psi_p^{(\tau)} = \Gamma^{nps} a_{nps}^{(\tau)} \eta$. Thus, the Ansatz (\ref{psif}) is traceless and
fulfills the Rarita-Schwinger equation provided that
\beq
a_{nps}^{(\tau)} \, \Gamma^{nps} \eta = 0
\label{b3cond}
\eeq
To prove that the $a^{(\tau)}$ satisfy this condition we have to rely on further properties of $G_2$ manifolds. 

It is known that the harmonic forms on $Y$ split according to the branching of $SO(7)$ 
representations into those of $G_2$ \cite{jbook}.
The Betti numbers reflect this decomposition. For instance, since a 3-form transforms in the \r{35} of $SO(7)$ it follows that
$b^3=b^3_{\r1} + b^3_{\r7} + b^3_{\r{27}}$, where the subscripts indicate the $G_2$ representation. Moreover, on $Y$
$b^3_{\r7}=0$ and $b^3_{\r1}=1$ corresponds to the invariant 3-form $\varphi$. The remaining 3-forms $a^{(\tau)}$ are
in the \r{27} and are characterized by  $a^{(\tau)} \wedge \varphi=0$ and $a^{(\tau)} \wedge \Phi=0$ \cite{Bryant}.
These two conditions in turn imply
\beq
a^{(\tau)}_{mnp} \Phi^{\ell mnp}=0 \quad ; \quad   a^{(\tau)}_{mnp} \varphi^{mnp}=0
\label{tfprop}
\eeq
Now, given (\ref{tfour}), it can be shown that \cite{House}
\beq
\Gamma^{mnp} \eta = - i\varphi^{mnp}\eta + \Phi^{mnp\ell} \Gamma_\ell \eta
\label{3geta}
\eeq
Contracting with $a^{(\tau)}_{mnp}$ and substituting (\ref{tfprop}) yields the desired result (\ref{b3cond}).

Harmonic 2-forms, $a^{(\upsilon)}_{pq}$, $\upsilon=1, \cdots, b^2$, give further zero modes
\beq
\Psi_p^{(\upsilon)} = a^{(\upsilon)}_{pq} \, \Gamma^{q} \eta
\label{psis}
\eeq
We now obtain $\Gamma^{mnp} D_n \Psi_p^{(\upsilon)} = -\frac12 g^{mr} D_r(\Gamma^{np} a_{np}^{(\upsilon)} \eta)$ and
$\Gamma^p \Psi_p^{(\upsilon)} = \Gamma^{np} a_{np}^{(\upsilon)} \eta$. Then, the condition
\beq
a_{np}^{(\upsilon)} \, \Gamma^{np} \eta = 0
\label{b2cond}
\eeq
guarantees that the $\Psi_p^{(\upsilon)}$ satisfy $\Gamma^p \Psi_p^{(\upsilon)}=0$, and the Rarita-Schwinger equation.
The number of harmonic 2-forms splits as $b^2=b^2_{\r7} + b^2_{\r{14}}$, and $b^2_{\r7}=0$ \cite{jbook}. Thus, the $a^{(\upsilon)}$ 
are in the \r{14} and are characterized by $a^{(\upsilon)}\wedge \varphi = -{}^*a^{(\upsilon)}$ \cite{Bryant}.
Taking dual gives $a^{(\upsilon)}_{mn} = -\frac12 a^{(\upsilon)\, pq} \Phi_{mnpq}$. Contracting with $\varphi^{mn\ell}$
we obtain 
\beq
a^{(\upsilon)}_{mn} \varphi^{mn\ell} = 0
\label{2fprop}
\eeq
by virtue of the identity $\Phi_{mnpq}\varphi^{mn\ell} = 4 \varphi^\ell_{pq}$ (see e.g. appendix B in \cite{House}).
The result (\ref{b2cond}) finally follows from $\Gamma^{np} \eta = - i\varphi^{np\ell} \Gamma_\ell \eta$.

In conclusion, the $(b^3+b^2)$ gravitino zero modes are given by (\ref{psi0}), (\ref{psif}), and (\ref{psis}).


{\small

\begin{thebibliography}{99}

\bibitem{j1}
D. Joyce,  
``Compact Riemannian 7-manifolds with holonomy $G_2$. I'', 
Journal of Differential Geometry 43 (1996), 291. 

\bibitem{j2}
D. Joyce, 
``Compact Riemannian 7-manifolds with holonomy $G_2$. II'', 
Journal of Differential Geometry 43 (1996), 329. 

\bibitem{jbook}
D. Joyce,  ``Compact manifolds with special holonomy'',
Oxford University Press, 2000. 

\bibitem{papadopoulos}
G.~Papadopoulos and P.~K.~Townsend,
``Compactification of D = 11 supergravity on spaces of exceptional
holonomy,'' Phys.\ Lett.\  B {\bf 357} (1995) 300
[arXiv:hep-th/9506150].

\bibitem{ag}
B.~S.~Acharya and S.~Gukov,
``M theory and Singularities of Exceptional Holonomy Manifolds,''
Phys.\ Rept.\  {\bf 392} (2004) 121
[arXiv:hep-th/0409191].

\bibitem{dhvw}
L.~J.~Dixon, J.~A.~Harvey, C.~Vafa and E.~Witten,
``Strings On Orbifolds,
'' Nucl.\ Phys.\  B {\bf 261}, 678 (1985);
``Strings On Orbifolds. 2,''
Nucl.\ Phys.\  B {\bf 274}, 285 (1986).

\bibitem{sv}
 S.~L.~Shatashvili and C.~Vafa,
 ``Superstrings and manifold of exceptional holonomy,''
 Selecta Math.\  {\bf 1} (1995) 347
 [arXiv:hep-th/9407025].

\bibitem{acharya1}
B.~S.~Acharya,
``N=1 M-theory-Heterotic Duality in Three Dimensions and Joyce Manifolds,'' arXiv:hep-th/9604133.

\bibitem{acharya}
 B.~S.~Acharya,
 ``Dirichlet Joyce manifolds, discrete torsion and duality,''
 Nucl.\ Phys.\  B {\bf 492} (1997) 591
 [arXiv:hep-th/9611036];
``On mirror symmetry for manifolds of exceptional holonomy,''
 Nucl.\ Phys.\  B {\bf 524} (1998) 269
 [arXiv:hep-th/9707186].

\bibitem{Majumder}
J.~Majumder,
``Type IIA orientifold limit of M-theory on compact Joyce 8-manifold of
Spin(7)-holonomy,''
JHEP {\bf 0201} (2002) 048
[arXiv:hep-th/0109076].

\bibitem{Ferretti}
G.~Ferretti, P.~Salomonson and D.~Tsimpis,
``D-brane probes on $G_2$ orbifolds,''
JHEP {\bf 0203} (2002) 004
[arXiv:hep-th/0111050].

\bibitem{gk}
 M.~R.~Gaberdiel and P.~Kaste,
 ``Generalised discrete torsion and mirror symmetry for $G_2$ manifolds,''
 JHEP {\bf 0408} (2004) 001
[arXiv:hep-th/0401125].

\bibitem{Barrett}
 A.~B.~Barrett and A.~Lukas,
 ``Classification and moduli Kaehler potentials of $G_2$ manifolds,''
 Phys.\ Rev.\  D {\bf 71} (2005) 046004
 [arXiv:hep-th/0411071].

\bibitem{Eguchi}
T.~Eguchi and Y.~Sugawara,
``String theory on G(2) manifolds based on Gepner construction,''
Nucl.\ Phys.\  B {\bf 630} (2002) 132
[arXiv:hep-th/0111012].

\bibitem{bb}
R.~Blumenhagen and V.~Braun,
``Superconformal field theories for compact $G_2$ manifolds,'' JHEP {\bf 0112} (2001) 006
[arXiv:hep-th/0110232].

\bibitem{sy}
K.~Sugiyama and S.~Yamaguchi,
``Cascade of special holonomy manifolds and heterotic string theory,''
Nucl.\ Phys.\  B {\bf 622} (2002) 3 [arXiv:hep-th/0108219].

\bibitem{DNP}
M.~J.~Duff, B.~E.~W.~Nilsson and C.~N.~Pope,
``Kaluza-Klein Supergravity,'' Phys.\ Rept.\  {\bf 130} (1986) 1.

\bibitem{gpp}
G.~W.~Gibbons, D.~N.~Page and C.~N.~Pope,
``Einstein Metrics on $S^3$, $R^3$ and $R^4$ Bundles,'' Commun.\ Math.\ Phys.\  {\bf 127} (1990) 529.

\bibitem{imnq}
L.~E.~Ib\'a\~nez, J.~Mas, H.~P.~Nilles and F.~Quevedo,
``Heterotic Strings In Symmetric And Asymmetric Orbifold Backgrounds,''
Nucl.\ Phys.\  B {\bf 301} (1988) 157.

\bibitem{pol}
J. Polchinski,  ``String Theory'', Vol. II.
Cambridge University Press, 1998. 

\bibitem{Hamidi}
S.~Hamidi and C.~Vafa,
``Interactions on Orbifolds,'' Nucl.\ Phys.\  B {\bf 279} (1987) 465.

\bibitem{House}
T.~House and A.~Micu,
``M-theory compactifications on manifolds with $G_2$ structure,''
Class.\ Quant.\ Grav.\  {\bf 22} (2005) 1709 [arXiv:hep-th/0412006].

\bibitem{chsw}
P.~Candelas, G.~T.~Horowitz, A.~Strominger and E.~Witten,
``Vacuum Configurations For Superstrings,''
Nucl.\ Phys.\  B {\bf 258} (1985) 46.

\bibitem{gam}
P.~Candelas and D.~J.~Raine,
``Compactification and supersymmetry in D = 11 supergravity,''
Nucl.\ Phys.\  B {\bf 248} (1984) 415; \\
W.~M\"uck,
``General (anti-)commutators of gamma matrices,'' arXiv:0711.1436 [hep-th].

\bibitem{Bryant}
R.~L.~Bryant,
``Some remarks on ${\rm G}_2$-structures,''
arXiv:math/0305124.


\end{thebibliography}
}
\end{document}